\documentclass[twoside,fleqn]{article}
\usepackage{latexsym,espcrc2,epsfig}

\title{
Improved actions for the two-dimensional $\sigma$-model}

\author{
  Sergio Caracciolo${}^{\rm a}$, 
  Andrea Montanari\address{Scuola Normale Superiore and INFN,
      Sezione di Pisa,
      I-56100 Pisa, ITALIA},
 $\!$and
  Andrea Pelissetto\address{Dipartimento di Fisica and INFN,
      Universit\`a degli Studi di Pisa,
      I-56100 Pisa, ITALIA}
}

\begin{document}

\begin{abstract}
For the $O(N)$ $\sigma$-model we studied the improvement program for 
actions with two- and four-spin interactions. An interesting example 
is an action which is reflection-positive, on-shell improved, and has 
all the coupling defined on an elementary plaquette.
We show the large $N$ solution and preliminary Monte Carlo results 
for $N=3$. 
\end{abstract}

\maketitle

\newcommand{\reff}[1]{(\ref{#1})}
\def\smfrac#1#2{{\textstyle\frac{#1}{#2}}}

\newcommand{\be}{\begin{equation}}
\newcommand{\ee}{\end{equation}}
\newcommand{\<}{\langle}
\renewcommand{\>}{\rangle}

\newcommand\FF{{\cal F}_\delta}

\def\spose#1{\hbox to 0pt{#1\hss}}
\def\ltapprox{\mathrel{\spose{\lower 3pt\hbox{$\mathchar"218$}}
 \raise 2.0pt\hbox{$\mathchar"13C$}}}
\def\gtapprox{\mathrel{\spose{\lower 3pt\hbox{$\mathchar"218$}}
 \raise 2.0pt\hbox{$\mathchar"13E$}}}

\def\bsigma{\mbox{\protect\boldmath $\sigma$}}
\def\bpi{\mbox{\protect\boldmath $\pi$}}
\def\btau{\mbox{\protect\boldmath $\tau$}}
\def\hatp{\hat p}
\def\hatl{\hat l}

\def\msbar{ {\overline{\hbox{\scriptsize MS}}} }
\def\normalmsbar{ {\overline{\hbox{\normalsize MS}}} }

\newcommand{\R}{\hbox{{\rm I}\kern-.2em\hbox{\rm R}}}

\def\smfrac#1#2{{\textstyle\frac{#1}{#2}}}

In recent years there has been much work in improving lattice actions
(see e.g. M. L\"uscher's and P. Hasenfratz' talks at this conference).
The idea behind all these attempts is to modify the lattice 
action with the addition of irrelevant operators in order to reduce 
lattice artifacts: the aim is to have scaling and finite-size-scaling 
at small values of the correlation length.

Two different approaches have been used. The original idea of 
Symanzik~\cite{Symanzik} consisted in adding, on the basis of power conting,
new operators to the action with coefficients such as to cancel corrections of 
order $O(g^{2n} a^2)$ in the 
correlation functions. It was later realized that it is possible to 
change the action so as to improve only on-shell quantities 
\cite{Luscher-Weisz}.  In this case the number of necessary operators is in 
general reduced. This approach can be applied 
order by order in perturbation theory in a straightforward 
(although computationally difficult) way. Recently the 
idea has been successfully implemented at a non-perturbative level
\cite{nonptimpr} for the fermionic actions (here the corrections are of 
order $a$), obtaining actions which are full improved 
at order $O(a)$.

A different approach is the perfect action program of Ref. 
\cite{perfect-action}: here the starting point is an action which is 
the fixed point of a renormalization-group transformation. The 
relation between the Symanzik approach and the fixed-point actions
has been recently clarified in Ref. \cite{no-one-loop-improved}. 
Fixed-point actions are {\em on-shell}, {\em tree-level} improved 
to {\em all orders in} $a^2$ (this last condition should not be very relevant
--- at least for the scaling of standard observables --- as quantum corrections
will introduce again terms of order $a^2$). Therefore these actions are 
particular Symanzik theories. Of course the main question is if this particular 
procedure chooses among all Symanzik-improved theories those which have 
a ``better" behaviour. We do not know the answer to this question, but 
we can try to understand it by comparing the behaviour of fixed-point
actions and other theories which are improved \`a la Symanzik using 
different criteria (for instance semplicity, locality .....). Here 
we will study this problem in the context of the two-dimensional 
$\sigma$-model focusing on the behaviour of the mass gap in a strip. 

Let us begin by discussing tree-level improvement and in particular 
the relation between
on-shell and off-shell improvement. In this case one can 
show that for generic actions which have only two-spin couplings
on-shell and off-shell improvement at tree level are identical.
More precisely, consider an action of the form
\be
S =\, \sum_{x,y} J(x-y) \bsigma_x \cdot \bsigma_y \;\; ,
\label{eq2}
\ee
such that:
\begin{enumerate}
\item the Fourier transform $\hat{J}(p)$  of $J(x)$ is continuous
   (locality);
\item $\hat{J}(p) - \hat{J}(0) = 0$ only for $p=(0,0)$ in the 
  Brillouin zone (correct continuum limit);
\item $\hat{J}(p) - \hat{J}(0) = - \alpha p^2 + O(p^4)$ for $p\to 0$
  with $\alpha > 0$ (ferromagnetic).
\end{enumerate}
Then tree-level improvement (both on-shell and off-shell) requires 
that, for $p\to 0$,
\be
\hat{J}(p) - \hat{J}(0) = - \alpha p^2 + O(p^6).
\ee
The classical example is the action proposed originally by Symanzik
\cite{Symanzik}
\be
S^{Sym} = 
    \sum_{x\mu} \left( {4\over3} \bsigma_x \cdot \bsigma_{x+\mu} \, -\, 
         {1\over12} \bsigma_x \cdot \bsigma_{x+2\mu} \right).
\label{SSymanzik}
\ee
It is possible to obtain actions which are on-shell, but not off-shell,
improved if one adds also four-spin couplings. Indeed, if 
$S^{cl}$ is the classical continuum action
\be
S^{cl} = {1\over2} \int d^2 x (\bsigma\cdot \Box\bsigma) ,
\ee
($\Box = \sum_\mu \partial_\mu^2$), then it is easy to check that
\begin{eqnarray}
\Box \bpi \cdot {\delta S^{cl}\over \delta \bpi} - 
(\bsigma\cdot \Box\bsigma) \bpi \cdot {\delta S^{cl}\over \delta \bpi} =
\nonumber \\
\Box \bsigma\cdot \Box\bsigma - (\bsigma\cdot \Box\bsigma)^2
\end{eqnarray}
Thus a lattice action whose continuum limit is 
\begin{eqnarray}
S &=& \int d^2 x \left[ {1\over2} (\bsigma\cdot \Box\bsigma) \right.
\nonumber \\
&+& \left.
  {A a^2\over2} (\bsigma\cdot \Box^2 \bsigma - 
    (\bsigma\cdot \Box \bsigma)^2 ) + O(a^4) \right]
\label{Honshellimp}
\end{eqnarray}
is on-shell improved for any $A$ since the added term vanishes 
when one uses the classical equations of motion. In a more rigorous way
one can show that, up to terms of order $O(a^4)$, a change of variables
allows to get rid of the term proportional to $A$ in 
(\ref{Honshellimp}). It is enough to reexpress the action in terms of 
\be
\bpi' = \bpi + {A a^2\over2} (\Box \bpi - \bpi (\bsigma\cdot \Box\bsigma)).
\ee

It is easy to write down an action with the continuum limit 
(\ref{Honshellimp}). A possible choice with all the couplings defined on a 
plaquette is given by \cite{CP-PLB}
\begin{eqnarray}
S^{pl} \!\!\! &\!\! =& \!\!\!\!\!
   \sum_{x} \left( {2\over3} \sum_\mu \bsigma_x \cdot \bsigma_{x+\mu}
    \, +\, {1\over6} \sum_{\hat{d}}\bsigma_x \cdot \bsigma_{x+\hat{d}} \right)
\nonumber \\
    &-& {1\over24} \sum_x \sum_{i=1}^4 
     \left(\sum_{{\mu_i}} 
    (\bsigma_x\cdot\bsigma_{x+{\mu_i}} -1) \right)^2
\label{Splaquette}
\end{eqnarray}
where $\hat{d}$ are the two diagonal vectors $(1,\pm1)$,
$\mu_1$ runs over the vectors $(1,0)$ and $(0,1)$,
$\mu_2$ over $(-1,0)$ and $(0,1)$,
$\mu_3$ over $(-1,0)$ and $(0,-1)$ and
$\mu_4$ over $(1,0)$ and $(0,-1)$ (the sum over $i$ symmetrizes 
over the four plaquettes stemming from the point $x$).

It is easy to verify the following properties of the action $S^{pl} $:
\begin{enumerate}
\item it is the most local action satisfying (\ref{Honshellimp});
\item it is reflection positive with respect to lattice planes;
\item the ordered configuration is the unique maximum of $S^{pl} $;
\item under Wolff's embedding only two-spin  couplings are 
generated, so that one can simulate $S^{pl}$ by a standard 
Wolff's algorithm.
\end{enumerate}
It is easy to study the large-$N$ limit of the various actions. The 
large-$N$ solution of (\ref{Splaquette}) can be obtained following 
the method used for mixed $O(N)$/$RP^{N-1}$ models 
\cite{Magnoli-Ravanini}. The two-point function is given by
\be
G(x) = \frac{1}{\beta(1+\omega)}\ 
\int {d^2p\over (2 \pi)^2} \frac{e^{ipx}}{{w}(p;\omega)+m^2},
\ee
where
\be
{w}(p;\omega)\equiv \hat{p}^2 -
{1\over 6} {\hat{p}_1^2 \hat{p}_2^2\over 1 + \omega};
\ee
here $\omega$ and $m$ are related to $\beta$ by the gap equations
\begin{eqnarray}
\beta (1+\omega)&=& \int {d^2p\over (2 \pi)^2}
             \frac{1}{{w}(p;\omega)+m^2}\; , \\
6\beta\omega(1+\omega) & =&\int {d^2p\over (2 \pi)^2}
        \frac{\hat{p}^2}{{w}(p;\omega)+m^2} \; ,
\end{eqnarray}
and $\hat{p} = 2 \sin(p/2)$.

\begin{figure}[t]
\vspace*{-4mm} \hspace*{-0cm}
\begin{center}
\epsfxsize = 0.45\textwidth
\leavevmode\epsffile{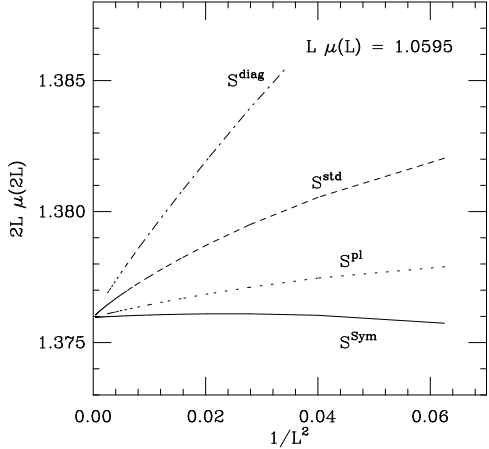}
\end{center}
\vspace*{-2cm}
\caption{
Corrections to FSS for the mass gap $\mu(L)$ for 
$N=\infty$ for the four actions described in the text}
\label{fig1}
\end{figure}

It is also possible to compute analytically the corrections to finite-size 
scaling. If one considers a strip $L\times \infty$ and the mass-gap 
$\mu(L)$, in the FSS limit one finds
\be
{\mu(L)\over \mu(\infty)} = 
 f(x) \left[ 1 + {g_0(x)\over L^2} + O(1/(L^2 \log L))\right]
\ee
where $x\equiv \mu(L) L$, and all functions can be analytically computed.
The result is similar to what has been found for general 
two-spin actions \cite{CPLat96}: tree-level improvement cancels 
corrections of order $\log L/L^2$. In the large-$N$ limit we can easily
compare the various actions. In Fig. \ref{fig1} we report 
$\mu(2L)2L$ for $x=1.0595$ for the Symanzik action, the plaquette
action, the standard action $S^{std}$ with nearest neighbour interactions
and the diagonal action $S^{diag}$ defined by \reff{Splaquette} 
without the four-spin term. 
Two observations are in order: first of all,
as expected, the non-improved actions show larger corrections to scaling
than their improved counterparts; less expected is instead the fact that 
the plaquette action 
is clearly worse than the standard Symanzik one in spite of the ``nicer"
theoretical properties (more local, reflection positive $\ldots$). 
Morever $S^{Sym}$ scores much better than expected: on this scale
no corrections are seen up to $L=5$.

\begin{figure}[t]
\vspace*{-4mm} \hspace*{-0cm}
\begin{center}
\epsfxsize = 0.45\textwidth
\leavevmode\epsffile{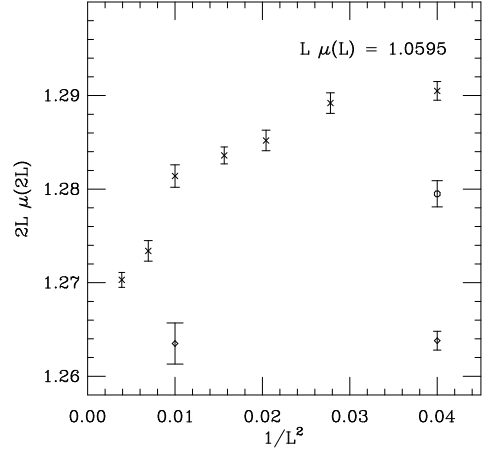}
\end{center}
\vspace*{-2cm}
\caption{
$\mu(2L) 2L$ for $N=3$ for fixed 
$L \mu(L)  = 1.0595$. The crosses refer to the standard action
(Ref. \protect\cite{LWW}), the diamonds to the perfect action
(Ref. \protect\cite{perfect-action}) and the circle to the 
plaquette action.}
\label{fig2}
\end{figure}

It is also interesting to perform the comparison of the various actions for 
$N=3$. In Fig. \ref{fig2} we report the same data as Fig. \ref{fig1}
but for
$N=3$ together the perfect-action results of Ref.
\cite{perfect-action}.  No data for the Symanzik action are yet
available. The results look very similar to those for $N=\infty$: in
particular the plaquette action  shows corrections which are only
half the size of the those of $S^{std}$  and it is indeed much worse
than the perfect action.

%
%
%


\begin{thebibliography}{9}

\bibitem{Symanzik} 
K. Symanzik, in {\em ``Mathematical problems in 
theoretical physics"}, R. Schrader et al. eds., (Springer, Berlin, 1982);
Nucl. Phys. {B226} (1983) 187;
{\em ibid.} 205.

\bibitem{Luscher-Weisz} 
L\"uscher and P. Weisz, Comm. Math. Phys. {97} (1985) 59;
{\em erratum} {98} (1985) 433.

\bibitem{nonptimpr} M. L\"uscher, S. Sint, R. Sommer, P. Weisz,
H. Wittig and U. Wolff, Nucl. Phys. B (Proc. Suppl.) 53 (1997) 905.

\bibitem{perfect-action} 
P. Hasenfratz and F. Niedermayer, 
Nucl. Phys. {B414} (1994) 785.

\bibitem{no-one-loop-improved} 
P. Hasenfratz and F. Niedermayer,
Fixed point actions in one-loop perturbation theory,
hep-lat/9706002.

\bibitem{CP-PLB}
A. Caracciolo and A. Pelissetto, Phys. Lett. {B402} (1997) 335.

\bibitem{Magnoli-Ravanini} N. Magnoli and F. Ravanini,
Z. Phys. C31 (1986) 567.

\bibitem{CPLat96} S. Caracciolo and A. Pelissetto, 
Nucl. Phys. B (Proc. Suppl.) 53 (1997) 693.

\bibitem{LWW} M. L\"uscher, P. Weisz and U. Wolff, 
Nucl. Phys. B359 (1991) 221.

\end{thebibliography}
\end{document}